\documentclass{article}

\usepackage{hyperref}
\usepackage[style=numeric-comp,sorting=none]{biblatex}
\usepackage{graphicx}
\usepackage{amsmath}
\usepackage{amssymb}
\usepackage{tensor}
\usepackage{url}
\usepackage{authblk}
\usepackage{tikz}

\newcommand{\sd}{\mathrm{d}}

\newcommand{\bra}[1]{\left<#1\right|}
\newcommand{\ket}[1]{\left|#1\right>}

\DeclareMathOperator{\sgn}{sgn}

\title{Unruh detectors, Feynman diagrams, acceleration and decay}
\author[1]{Wim Beenakker}
\author[1]{David Venhoek}
\affil[1]{Institute for Mathematics, Astrophysics and Particle Physics, Radboud University
Nijmegen, Heyendaalseweg 135, Nijmegen, The Netherlands}
\date{December 1, 2025}

\allowdisplaybreaks

\begin{document}

\maketitle

\begin{abstract}
We present a method for relating the transition rate of an accelerated Unruh-deWitt detector to the rate of the same detector when stationary in Minkowski space. Furthermore, we show that when using the detector as a model for decay, its transition rate can be related directly to the decay rate obtained from QFT. Combined this provides a straightforward method for calculating the decay rate of accelerated particles to first order in the coupling constants.
\end{abstract}

Unruh-deWitt detectors are an abstract model for particle detectors. They were first devised for the study of the Unruh effect and Hawking radiation \cite{Unruh, DeWitt}, and have since been used to study the observability of the Unruh effect, both directly \cite{bell1,bell2} and indirectly through processes such as decay \cite{Muller, Matsas, self_decay}, cooling \cite{Garay}, or entanglement \cite{Zhou}.

Beyond the study of the Unruh effect, Unruh-deWitt detectors have also seen application in the more general study of entanglement (see e.g. \cite{VALENTINI1991321,salton2015acceleration,PhysRevD.97.125002,PhysRevD.105.065016,agullo2024multimodenaturespacetimeentanglement,ribesmetidieri2024inflationdoescreateentanglement}), and to study the particle contents of more general curved spacetimes, in for example \cite{PhysRevD.13.2188, PhysRevD.15.2738, Louko_2008}. This broad range of applications has also led to efforts to study the theoretical underpinnings of Unruh-deWitt detectors, and how they relate to other notions of measurements, such as those found in algebraic quantum field theory and quantum measurement theory \cite{Fewster2020}.

In this paper we will approach the Unruh-deWitt detector as a model for decay, concentrating on its behaviour in flat spacetime. After setting up our framework and conventions, we will show in section~\ref{sec:accelerated_decay} that the calculation of the transition rate of an accelerated Unruh-deWitt detector can be reduced to calculating an integral over the decay rate as a function of energy difference between the levels of the same detector in a non-accelerated setting.

After going over the main QFT results we will need and the conventions we use for that, section~\ref{sec:equivalence} contains a proof of a result that shows that this decay rate can be calculated in a straightforward manner through Feynman diagrams and phase space integrals. This completes a framework for calculating the decay rate of accelerated particles to first order in the coupling constants in a straightforward, mechanical manner.

Finally, section~\ref{sec:example} provides a short example to give a flavor of how these results apply to a concrete calculation. This example is also used to explicitly work out a number of checks on our results.

\section{Conventions}

We try, where possible, to follow the conventions used in \cite{PeskinSchroeder}. In particular, we use natural units, where $\hbar = c = 1$. Where concrete numbers are needed for masses and energy scales, one mass or energy is taken to be arbitrary, providing a scale, and all other masses/energies are expressed relative to this scale mass. For the metric, we use the mostly minus convention $(+,-,-,-)$.

Throughout this work we will be using a number of special mathematical functions. The function $\sgn(x)$ will denote the sign of $x$, being $-1$ if $x$ is below zero, $+1$ if $x$ is above zero, and $0$ if $x$ equals zero. The function $K_a(b)$ is the modified Bessel function of the second kind. The function $\Gamma(z)$ represents the Gamma function, and we will use $B(z_1, z_2) = \frac{\Gamma(z_1)\Gamma(z_2)}{\Gamma(z_1+z_2)}$ to denote the Beta function. For all of these, we will use the rigorous definition from \cite{DLMF}, which matches common usage. The K\"all\'en function $\lambda(x,y,z) = x^2+y^2+z^2-2xy-2xz-2yz$ will be used to simplify several expressions in kinematics.

In a number of places, we deal with both the energy, the momentum as well as the energy-momentum vector of a particle. When integrating or otherwise dealing with a spatial momentum $\vec{p_x}$, unless otherwise stated the quantity $E_x$ is the corresponding on-shell energy for the particle with that momentum, and $p_x$ denotes the associated energy-momentum vector.

When dealing with momentum conservation in QFT, we will occasionally have need to denote the energy-momentum vector of a massive particle in its rest frame. For this, we overload notation on the mass of that particle $m_p$. When such a mass occurs in a position where an energy-momentum vector is expected, it will denote the rest-frame energy-momentum vector $(m_p, 0,\ldots,0)$.

\section{Unruh detectors as model for decay}\label{sec:unruhmodel}

We consider an Unruh detector model that is a direct generalisation of the model used in \cite{self_decay}, taking the detector to be a two level system with an energy difference $\Delta$ between the states. This is coupled to $n$ otherwise non-interacting real scalar fields. We denote the masses of these fields with $m_i$, which may be zero for some or all fields. Although the Unruh detector as a model can be used in arbitrary spacetimes, we will here restrict to the case of flat spacetime, or in other words, a spacetime with the Minkowski metric $\eta\indices{^\mu^\nu}$. We will also ignore any alternate topologies for this spacetime, focusing only on the unbounded plane.

The interaction term is taken to be $GM\prod_i \phi_i(x(\tau))$. Here $G$ is the strength of the coupling, and the $\phi_i$ are the field operators for each of the $n$ massive fields. $M$ is a time independent dimensionless operator on the two level system encoding how it interacts with the fields. It will typically be of the form $\ket{0}\bra{1}+\ket{1}\bra{0}$. The entire interaction is assumed to take place only along some path $x(\tau)$, which is assumed to be timelike and parameterized by its length. This path $x(\tau)$ will be interpreted as the path that the detector follows in spacetime.

These components give the following action for the system:
\begin{align}
S &= \int \sd^{d} x\sum_i\frac{1}{2}\left(\eta\indices{^\mu^\nu}\partial\indices{_\mu}\phi_i\partial\indices{_\nu}\phi_i-m_i^2\phi_i^2\right)\nonumber\\*
&\phantom{=}\quad + \int\sd\tau\left(\Delta\ket{1}\bra{1}+GM\prod_i\phi_i(x(\tau))\right).
\end{align}
In the interaction picture, this gives a time evolution of the states of
\begin{align}
\frac{\partial}{\partial\tau}\ket{s} &= -iGM(\tau)\prod_i\phi_i(x(\tau))\ket{s},
\end{align}
where
\begin{align}
M(\tau) &= e^{i\Delta\tau\ket{1}\bra{1}}Me^{-i\Delta\tau\ket{1}\bra{1}}.
\end{align}

We will consider the rate of transition from the detector's $\ket{1}$ state to the $\ket{0}$ state, with the fields starting in the non-interacting vacuum state $\ket{\psi_0}$ and ending in some arbitrary state $\ket{\psi}$. To derive this rate we start with the probability of transition between these states over some time interval $2T$ to lowest order in $g$
\begin{align}
P_{10}(2T) &= \sum_{\ket{\psi}}\left|\bra{0}\otimes \bra{\psi}\int\limits_0^{2T}\sd \tau GM(\tau)\prod_i\phi_i(x(\tau))\ket{1}\otimes\ket{\psi_0}\right|^2\\
&= G^2\left|\bra{0}M\ket{1}\right|^2\int\limits_0^{2T}\sd\tau_1\int\limits_0^{2T}\sd\tau_2 e^{i\Delta(\tau_1-\tau_2)}\nonumber\\*
&\phantom{=}\quad\quad\quad\quad\quad\quad\quad\quad \cdot \bra{\psi_0}\prod_i\phi_i(x(\tau_1))\phi_i(x(\tau_2))\ket{\psi_0}.
\end{align}
The sum is taken over all states for the combined fields, which enables using the completeness relation in the second equality. As these are non-interacting fields, this can be further simplified by noting that the expectation value of the combined interference term is just the product of that for the individual fields, yielding
\begin{align}
P_{10}(2T) &= G^2\left|\bra{0}M\ket{1}\right|^2\int\limits_0^{2T}\sd\tau_1\int\limits_0^{2T}\sd\tau_2 e^{i\Delta(\tau_1-\tau_2)}\nonumber\\*
&\phantom{=}\quad\quad\quad\quad\quad\quad\quad\quad\quad\quad\quad \cdot \prod_i\bra{\psi_{i,0}}\phi_i(x(\tau_1))\phi_i(x(\tau_2))\ket{\psi_{i,0}}
\end{align}
with the $\psi_{i,0}$ the vacuum states for the individual $n$ fields.

We can translate this to a rate by dividing by $2T$ and taking the infinite time limit:
\begin{align}\label{eq:raw_decay_rate}
\Gamma &= \lim_{T\rightarrow\infty}\frac{G^2}{2T}\left|\bra{0}M\ket{1}\right|^2\int\limits_0^{2T}\sd\tau_1\int\limits_0^{2T}\sd\tau_2 e^{i\Delta(\tau_1-\tau_2)}\nonumber\\*
&\phantom{=} \quad\quad\quad\quad\quad\quad\quad\quad\quad\quad\quad\quad \cdot \prod_i\bra{\psi_{i,0}}\phi_i(x(\tau_1))\phi_i(x(\tau_2))\ket{\psi_{i,0}}
\end{align}

To view this as a model for decay we will identify the various components of the Unruh detector model with the particles involved in the decay process. For this, we will use the same approach as in \cite{self_decay}, mapping the decaying particle to the $\ket{1}$ state of the detector, and the heaviest decay product to the $\ket{0}$ state. We take the energy difference $\Delta$ of the detector to be the mass difference between the decaying particle and the heaviest decay product. All other decay products are treated as fields, and we assume these to interact with the detector through a single interaction vertex.

In the mapping of both the decaying particle and its heaviest decay product to the detector, we assume that the trajectory of the heaviest decay product after decay matches that of the original decaying particle. This corresponds to taking a limit towards infinity of the masses of both these particles, whilst keeping the mass difference $\Delta$ between them constant.

To make this more concrete, let us look at the alpha decay of $^{210}\text{Po}$. We map the decaying particle and its heaviest decay product to the detector, which in this case will mean mapping the $^{210}\text{Po}$ nucleus to the $\ket{1}$ state, and the decayed nucleus $^{206}\text{Pb}$ to the $\ket{0}$ state. The alpha particle is then treated as the particle belonging to the field $\phi_1$. Thus, a single decay of $^{210}\text{Po}$ corresponds with a single transition of the Unruh detector from the $\ket{1}$ state to the $\ket{0}$ state, producing one particle belonging to the $\phi_1$ field.

\section{Relations between accelerated and non-accelerated decay}\label{sec:accelerated_decay}

The above has provided us with an interpretation of Unruh detectors as a model for decay. This mapping enables us to apply the primary advantage of the Unruh detector, i.e. allowing the studying of interactions along arbitrary paths, to the process of decay. In particular, this will allow us to study how an external force accelerating the decaying particle influences the decay of that particle.

To do this, we will consider two classes of trajectories for $x(\tau)$. First, we consider inertial detectors, which upon using Lorentz invariance can be described by a trajectory of the form $x(\tau) = (\tau, 0, \ldots, 0)$. The second class consist of detectors under uniform linear acceleration $a$, which corresponds to trajectories of the form $x(\tau) = (\frac{1}{a}\sinh(a\tau), \frac{1}{a}\cosh(a\tau), 0, \ldots, 0)$. We choose these paths with different $\tau=0$ origins for convenience in calculation.

Both of these classes of trajectories have the special property that they are \emph{self-similar}. With this, we mean that when shifting the eigentime parameterization, there is a matching Poincar\'e transformation ensuring that the shifted trajectory looks identical in the new coordinates to the unshifted trajectory in the old coordinates. In particular, this ensures that the distance between points $x(\tau_1)$ and $x(\tau_2)$ only depends on the difference $\tau_2 - \tau_1\equiv\kappa$.

Furthermore, Poincar\'e invariance of the theory ensures that the 2-point correlator $\bra{\psi_{i,0}}\phi_i(x(\tau_1))\phi_i(x(\tau_2))\ket{\psi_{i,0}}$ is merely a function of the (timelike) distance between $x(\tau_1)$ and $x(\tau_2)$. Therefore, we will define the function $g_i(t) = \bra{\psi_{i,0}}\phi_i((0, 0, \ldots, 0))\phi_i((t, 0, \ldots, 0))\ket{\psi_{i,0}}$, and $\Delta_\tau(\kappa)$ as the distance between $x(\tau)$ and $x(\tau+\kappa)$ (where the choice of $\tau$ no longer matters due to self-similarity). This allows us to rewrite the decay rate to
\begin{align}
\Gamma &= \lim_{T\rightarrow\infty}\frac{G^2}{2T}\left|\bra{0}M\ket{1}\right|^2\int\limits_{-2T}^{2T}\sd\kappa \left(2T-\left|\kappa\right|\right) e^{-i\Delta\kappa}\prod_ig_i(\Delta_\tau(\kappa)).
\end{align}

Before moving on, it will be insightful to explicitly calculate $\Delta_\tau$ for the two classes of trajectories we consider. For the non-accelerated detectors, it is immediately clear that $\Delta_\tau(\kappa) = \kappa$. The situation is a bit more complicated for the accelerated detectors, but a straightforward calculation shows that for a detector accelerated with acceleration $a$, $\Delta_\tau(\kappa) = \frac{2}{a}\sinh\left(\frac{a\kappa}{2}\right)$.

Assuming that $\prod_i g_i(t)$ is sufficiently well-behaved, this further simplifies (see \cite{Titchmarsh} section 1.15 for the reasoning) to
\begin{align}\label{eq:decay_unruh}
\Gamma &= G^2\left|\bra{0}M\ket{1}\right|^2\int\limits_{-\infty}^\infty\sd\kappa e^{-i\Delta\kappa}\prod_ig_i(\Delta_\tau(\kappa)),
\end{align}
which is the Fourier transform of the function $\prod_ig_i(\Delta_\tau(\kappa))$. Note that if we define $g(t) = \prod_i g_i(t)$, this can be rewritten as the function composition $g(\Delta_\tau(\kappa))$.

This latter problem, the calculation of the Fourier transform of a composed function, has been studied in \cite{self_fourier}. We can use this to transform the calculation for a detector accelerated at rate $a$ to
\begin{align}\label{eq:apply_accel}
\frac{\Gamma}{G^2\left|\bra{0}M\ket{1}\right|^2} &= \frac{1}{2\pi}\int\limits_{-\infty}^{\infty}\sd\Delta'\left(\int\limits_{-\infty}^{\infty}\sd \kappa e^{-i\Delta' \kappa} g(\kappa)\right)H_a(\Delta, \Delta')
\end{align}
with
\begin{align}
H_a &= \int\limits_{-\infty}^{\infty}\sd \kappa e^{i\left(\Delta t - \frac{2\Delta'}{a}\sinh\left(\frac{a t}{2}\right)\right)}\\
&= \frac{4}{a}\int\limits_0^\infty\sd u\cos\left(\frac{2\Delta}{a} u\right)\cos\left(\frac{2\Delta'}{a}\sinh(u)\right)\nonumber\\*
&\phantom{=} + \frac{4}{a}\int\limits_0^\infty\sd u\sin\left(\frac{2\Delta}{a} u\right)\sin\left(\frac{2\Delta'}{a}\sinh(u)\right)\\
&= \frac{4}{a}e^{\sgn(\Delta')\frac{\pi \Delta}{a}}K_{\frac{2i\Delta}{a}}\left(\left|\frac{2\Delta'}{a}\right|\right).
\end{align}
The calculation of the explicit form of $H_a$ follows from formula 10.32.7 in \cite{DLMF}.

Note that in the above expression, the inner integral in the decay rate actually is just the non-accelerating decay rate, albeit with a different energy gap $\Delta'$ in the detector. Furthermore, the function $H_a$ is actually rapidly decreasing as $\Delta'$ goes to infinity, and numerically reasonably well behaved away from $\Delta' = 0$. This means that, as long as we can determine the non-accelerated decay rate of our model, the accelerated decay rate can be found with a relatively straightforward numerical integration.

This leaves open the question of whether for the physical processes we consider here, the functions $g$ and $\Delta_\tau$ are sufficiently well behaved. For the accelerated class of detectors, the form of the function $\Delta_\tau$ is actually one of the examples used in \cite{self_fourier}, and we won't repeat verification here. For $g$ however, it is significantly less easy to argue that it is square integrable. We will leave the details for appendix~\ref{ap:square_integrable}, but it turns out that when doing dimensional regularization, and assuming all particles involved have non-zero mass, there is indeed a range of dimensions $d$ for which this is the case. From this range, we can then analytically extend back to the dimension of interest. We will postulate that this procedure provides a valid result.

\section{Decay processes in ordinary QFT}

The results from the previous section imply that, once we can calculate the decay for a non-accelerated detector, we can find the same result in the accelerated case via a straightforward procedure. Unfortunately, calculating the decay rate of an Unruh detector can be quite challenging even in the non-accelerated case, due to the infinities in the two-point function and the need to regularize these. However, ordinary QFT provides a means of calculating the decay rate of particles that, at least to first order in the coupling constants, is much more straightforward to calculate.

We once again limit ourselves to the case of distinct decay products, considering only contact interactions. With those restrictions, the total decay rate from QFT is given by the expression
\begin{align}\label{eq:qftdecay}
\Gamma &= \frac{1}{2m_a}\int \frac{\sd^{d-1} \vec{p}_1}{\left(2\pi\right)^{d-1}}\frac{1}{2E_1}\cdots\int \frac{\sd^{d-1} \vec{p}_n}{\left(2\pi\right)^{d-1}}\frac{1}{2E_n}\nonumber\\*
&\phantom{=}\quad\quad\quad\quad\quad\quad\quad\quad \cdot \left|\mathcal{M}\left(m_a\rightarrow\left\{p_i\right\}\right)\right|^2\left(2\pi\right)^d\delta^d\left(p_a - \sum_i p_i\right)
\end{align}
where $m_a$ and $p_a$ are the mass and energy-momentum of the decaying particle, and the $p_i$ and $E_i$ the energy-momentum vectors and energies of the decay products. The matrix element $\mathcal{M}\left(m_a\rightarrow\left\{p_i\right\}\right)$ is usually calculated through Feynman diagrams. For the simple theories with a single contact interaction we are considering here, to first order in the coupling constant there is only a single diagram to consider:
\begin{center}
\begin{tikzpicture}
\draw (-2,0) node[anchor=east] {$p_a$} -- (0,0);
\draw (0,0) -- (2,1.5) node[anchor=west] {$p_1$};
\draw (0,0) -- (2,1.0) node[anchor=west] {$p_2$};
\filldraw (1.5, 0.3125) circle (1pt);
\filldraw (1.5, -0.1875) circle (1pt);
\filldraw (1.5, -0.6875) circle (1pt);
\draw (0,0) -- (2,-1.5) node[anchor=west] {$p_n$};
\end{tikzpicture}
\end{center}
This diagram gives a value of the matrix element of $-ig$.

Deriving and motivating this recipe for calculating decay is a significant result in QFT, requiring most of the major technical results in the field. An overview of it can be found in most textbooks on quantum field theory, such as \cite{PeskinSchroeder}. However, it will be useful to recall that, although usually calculated through Feynman diagrams, the matrix element is defined as
\begin{align}\label{eq:matrixel}
\lim_{T\rightarrow\infty} \bra{p_1\cdots p_n}e^{-2iHT} - 1\ket{p_a} &= i\mathcal{M}\left(p_a\rightarrow\left\{p_i\right\}\right)\left(2\pi\right)^d\delta^d\left(p_a-\sum_i p_i\right)
\end{align}
which, for the process considered here, is to first order equivalent to
\begin{align}
&-ig\bra{p_1\cdots p_n}_0\int\sd x \phi_{a,0}(x)\phi_{1,0}(x)\cdots\phi_{n,0}(x)\ket{p_a}_0\nonumber\\*
&\quad\quad\quad\quad\quad\quad\quad\quad\approx i\mathcal{M}\left(p_a\rightarrow\left\{p_i\right\}\right)\left(2\pi\right)^d\delta^d\left(p_a-\sum_i p_i\right)
\end{align}
where in the latter the fields and state operators are free-theory ones. Here $\ket{p_a}_0$ is the state with a single particle from the $\phi_a$ field with momentum $p_a$, and $\bra{p_1\cdots p_n}_0$ is the state with a single particle from field $\phi_1$ with momentum $p_1$, a single particle from field $\phi_2$ with momentum $p_2$ and so forth. Although we here used separate notation for the non-interacting fields, throughout the rest of the paper all fields and states will be taken to be of the non-interacting variety.

Especially for a single point interaction, and to lowest order in the coupling constants, Equations~\ref{eq:qftdecay} and~\ref{eq:matrixel} are reasonably straightforward to evaluate. More importantly, Equation~\ref{eq:qftdecay} is finite at lowest order, so even if it is hard to evaluate fully analytically, it is not unreasonable to expect it to be doable numerically. However, we derived the results of the previous section for Unruh detectors. Although they represent the same underlying physical process, it is a priori not clear that both the Unruh detector and QFT approach should yield the same decay rate.

\section{Equivalence between the Unruh detector model and ordinary QFT}\label{sec:equivalence}

It turns out that the two models really are equivalent, at least to lowest order in their respective coupling constants. We will show that here for real scalar fields, considering a decay from a particle of field $\phi_a$ into distinct particles from fields $\phi_1$ through $\phi_n$ through a point interaction. For our argument, we start with the quantity
\begin{align}\label{eq:X}
X &= \lim_{T\rightarrow\infty}\frac{1}{2T}\int \frac{\sd^{d-1}\vec{p_1}}{\left(2\pi\right)^{d-1}}\frac{1}{2E_1}\cdots\int \frac{\sd^{d-1}\vec{p_n}}{\left(2\pi\right)^{d-1}}\frac{1}{2E_n}\nonumber\\*
&\phantom{=}\quad\quad\quad\quad\quad\quad\quad\cdot \left|\bra{p_1\cdots p_n}\int\limits_{-T}^{T}\sd t\int\sd^{d-1}\vec{x} \phi_{a}(x)\phi_{1}(x)\cdots\phi_{n}(x)\ket{i}\right|^2
\end{align}
and show that, in the appropriate limits, $X$ is proportional to both the decay rate from the Unruh detector point of view, as well as the decay rate evaluated through QFT directly.

In the above expression, $\ket{i}$ denotes the initial state of the system before decay, or in other words, the decaying particle staying at the origin of the coordinate system. $\phi_a$ is the field of the particle that decays, and $\phi_i$ for $i$ between $1$ and $n$ are the fields of the decay products. Again the state $\bra{p_1\cdots p_n}$ is the state with a single particle from field $\phi_1$ with momentum $p_1$, a single particle from field $\phi_2$ with momentum $p_2$ and so forth. In the inner integral, we define $x = (t, \vec{x})$. All field operators and states are taken to be from the non-interacting theory. Finally, we will restrict ourselves to a situation where the difference between the mass of the decaying particle and that of the heaviest decay product is small compared to the mass of the decaying particle, e.g. $m_a - m_1 << m_a$.

\subsection{Reduction to Unruh detector}

We begin our proof that $X$ is proportional to the Unruh detector decay rate by using the fact that all the fields are non-interacting and distinct. This allows us to split the bracket expression in the integral in separate parts for each of the fields:
\begin{align}
X &= \lim_{T\rightarrow\infty}\frac{1}{2T}\int \frac{\sd^{d-1}\vec{p_1}}{\left(2\pi\right)^{d-1}}\frac{1}{2E_1}\int \frac{\sd^{d-1}\vec{p_2}}{\left(2\pi\right)^{d-1}}\frac{1}{2E_2}\cdots\int \frac{\sd^{d-1}\vec{p_n}}{\left(2\pi\right)^{d-1}}\frac{1}{2E_n}\nonumber\\*
&\phantom{=} \cdot\Bigg|\int\limits_{-T}^{T}\sd t\int\sd^{d-1}\vec{x} \bra{\psi_{a,0}}\phi_{a}(x)\ket{i}\bra{p_1}\phi_{1}(x)\ket{\psi_{1,0}}\bra{p_2}\phi_{2}(x)\ket{\psi_{2,0}}\nonumber\\
&\phantom{=}\quad\quad\cdots\bra{p_n}\phi_{n}(x)\ket{\psi_{n,0}}\Bigg|^2
\end{align}

Next, we will use the assumption that the state $\ket{i}$ represents a $\phi_a$ particle traveling on the timelike path of a detector at rest $x(t) = (t,0,\ldots,0)$. In particular, we assume that the wave packet $\bra{\psi_{a,0}}\phi_a(x)\ket{i}$ is sharply peaked around $x(t)$, much sharper than the de Broglie wavelength of the light particles. Using this, we can approximate:
\begin{align}
X &\approx \lim_{T\rightarrow\infty}\frac{1}{2T}\int \frac{\sd^{d-1}\vec{p_1}}{\left(2\pi\right)^{d-1}}\frac{1}{2E_1}\int \frac{\sd^{d-1}\vec{p_2}}{\left(2\pi\right)^{d-1}}\frac{1}{2E_2}\cdots\int \frac{\sd^{d-1}\vec{p_n}}{\left(2\pi\right)^{d-1}}\frac{1}{2E_n}\nonumber\\*
&\phantom{=}\cdot \Bigg|\int\limits_{-T}^{T}\sd t\int\sd^{d-1}\vec{x} \bra{\psi_{a,0}}\phi_{a}(x)\ket{i}\bra{p_1}\phi_{1}(x)\ket{\psi_{1,0}}\bra{p_2}\phi_{2}(x(t))\ket{\psi_{2,0}}\nonumber\\
&\phantom{=}\quad\quad\cdots\bra{p_n}\phi_{n}(x(t))\ket{\psi_{n,0}}\Bigg|^2
\end{align}

Finally, we will use the assumption that $\phi_a$ is approximately at rest and stationary, and that $m_1$ is close to $m_a$. This implies that both the $\phi_a$ particle, as well as the $\phi_1$ particle in the final state are approximately at rest. Thus, their contributions after integrating over the spatial components and the momentum of $\phi_1$ will be proportional to $e^{-im_at}$ for the $\bra{\psi_{a,0}}\phi_{a}(x)\ket{i}$ term, and to $e^{im_1t}$ for the $\bra{p_1}\phi_{1}(x)\ket{\psi_{1,0}}$ term, with an overall constant $C$ accounting for the precise shape of the initial state.

The previous approximation at first glance looks somewhat questionable, as the Heisenberg uncertainty principle requires a minimum amount of uncertainty on the momentum of $\phi_a$ given its earlier localisation. However, as both $\phi_a$ and $\phi_1$ are much more massive than the light particles, a center-of-mass momentum of $\phi_a$ in the decay mostly ends up as a momentum contribution to $\phi_1$, with only very small contributions to the light particles, compared to the maximum momentum they can get from decay at rest. This ensures both approximations can indeed be made at the same time.

Incorporating these approximations for the $\phi_1$ and $\phi_a$ particles into the above expression simplifies the decay rate to
\begin{align}
X &\approx \lim_{T\rightarrow\infty}\frac{1}{2T}\int \frac{\sd^{d-1}\vec{p_2}}{\left(2\pi\right)^{d-1}}\frac{1}{2E_2}\cdots\int \frac{\sd^{d-1}\vec{p_n}}{\left(2\pi\right)^{d-1}}\frac{1}{2E_n}\nonumber\\*
&\phantom{=}\quad \left|\int\limits_{-T}^{T}\sd t Ce^{-im_at}e^{im_1t}\bra{p_2}\phi_{2}(x(t))\ket{\psi_{2,0}}\cdots\bra{p_n}\phi_{n}(x(t))\ket{\psi_{n,0}}\right|^2
\end{align}

We can now use completeness, while at the same time defining $\Delta = m_a - m_1$, to obtain
\begin{align}
X &\approx \lim_{T\rightarrow\infty}\frac{\left|C\right|^2}{2T}\int\limits_{-T}^T\sd t_1\int\limits_{-T}^T\sd t_2 e^{-i\Delta(t_2-t_1)}\nonumber\\*
&\phantom{=}\quad\quad\quad\quad\quad\quad\quad\quad \cdot \prod_{i>1}\bra{\psi_{i,0}}\phi_i(x(t_1))\phi_i(x(t_2))\ket{\psi_{i,0}}.
\end{align}

This implies a proportionality between $X$ and the Unruh detector decay rate: $X \approx \frac{\left|C\right|^2}{G^2\left|\bra{0}M\ket{1}\right|^2}\Gamma_{\text{Unruh}}$.

\subsection{Reduction to QFT}

Let us now switch focus to showing $X$ to be proportional to the QFT decay rate. To start this process, we first note that the initial state $\ket{i}$ is by definition a single-particle state. Hence, there exists a wave packet $\phi_I(\vec{p_I})$ such that $\ket{i} = \int\frac{\sd^{d-1}\vec{p_I}}{(2\pi)^{d-1}}\frac{\phi_I(\vec{p_I})}{\sqrt{2E_I}}\ket{p_I}$. Note that $\phi_I$ is not the same as the field operator $\phi_i$, but rather a map from momentum to the complex numbers. This gives
\begin{align}
X &= \lim_{T\rightarrow\infty}\frac{1}{2T}\int \frac{\sd^{d-1}\vec{p_1}}{\left(2\pi\right)^{d-1}}\frac{1}{2E_1}\cdots\int \frac{\sd^{d-1}\vec{p_n}}{\left(2\pi\right)^{d-1}}\frac{1}{2E_n}\nonumber\\*
&\!\!\cdot \left|\int\frac{\sd^{d-1}\vec{p_I}}{(2\pi)^{d-1}}\frac{\phi_I(\vec{p_I})}{\sqrt{2E_I}}\bra{p_1\cdots p_n}\int\limits_{-T}^{T}\sd t\int\sd^{d-1}\vec{x} \phi_{a}(x)\phi_{1}(x)\cdots\phi_{n}(x)\ket{p_I}\right|^2
\end{align}

Next, we would like to use the definition of matrix elements to introduce them into the equation. However, this requires integration over the entire spacetime, which we only have in the limit $T\rightarrow \infty$. However, we cannot take this limit immediately. To deal with this issue, we define a function $\delta_T$ which in the limit becomes the delta function, but which encodes the approximation to the delta function that occurs when $T$ is still finite. Using this notation we find
\begin{align}
X &= \lim_{T\rightarrow\infty}\frac{1}{2T}\int \frac{\sd^{d-1}\vec{p_1}}{\left(2\pi\right)^{d-1}}\frac{1}{2E_1}\cdots\int \frac{\sd^{d-1}\vec{p_n}}{\left(2\pi\right)^{d-1}}\frac{1}{2E_n}\nonumber\\*
&\phantom{=}\quad \Bigg|\int\frac{\sd^{d-1}\vec{p_I}}{(2\pi)^{d-1}}\frac{\phi_I(\vec{p_I})}{\sqrt{2E_I}}\frac{1}{g}\mathcal{M}\left(p_I\rightarrow p_1\cdots p_n\right)\nonumber\\*
&\phantom{=}\quad\quad\cdot \left(2\pi\right)^d\delta^{d-1}\left(\vec{p_I}-\sum_i\vec{p_i}\right)\delta_T\left(E_I-\sum_i E_i\right)\Bigg|^2
\end{align}
The function $\delta_T$ can be made precise with ideas from analysis. However we only need two properties: that $\delta_T\rightarrow \delta$ as $T\rightarrow \infty$, and that $\lim_{T\rightarrow\infty}\frac{2\pi\delta_T(0)}{2T} = 1$.

Subsequently, we can write out the squared norm and use one of the sets of momentum delta functions to integrate out one of the two occurrences of $\vec{p_I}$.
\begin{align}
X &= \lim_{T\rightarrow\infty}\frac{1}{2T}\int \frac{\sd^{d-1}\vec{p_1}}{\left(2\pi\right)^{d-1}}\frac{1}{2E_1}\cdots\int \frac{\sd^{d-1}\vec{p_n}}{\left(2\pi\right)^{d-1}}\frac{1}{2E_n}\int\frac{\sd^{d-1}\vec{p_I}}{(2\pi)^{d-1}}\frac{\left|\phi_I(\vec{p_I})\right|^2}{2E_I}\nonumber\\*
&\phantom{=}\quad \cdot\frac{1}{g^2}\left|\mathcal{M}\left(p_I\rightarrow p_1\cdots p_n\right)\right|^2(2\pi)^d\delta^{d-1}\left(\vec{p_I}-\sum_i\vec{p_i}\right)\nonumber\\*
&\phantom{=}\quad\cdot\delta_T\left(E_I-\sum_i E_i\right)(2\pi)\delta_T\left(E_I-\sum_i E_i\right)
\end{align}

Note that the energy conservation delta function effectively occurs twice (even though it is still a $\delta_T$ at this point). Thus, we can replace its argument with $0$ in the second occurrence, and use that delta function to cancel the division by $2T$. This allows us to finally take the $T$ limit, obtaining
\begin{align}
X &= \frac{1}{g^2}\int \frac{\sd^{d-1}\vec{p_1}}{\left(2\pi\right)^{d-1}}\frac{1}{2E_1}\cdots\int \frac{\sd^{d-1}\vec{p_n}}{\left(2\pi\right)^{d-1}}\frac{1}{2E_n}\int\frac{\sd^{d-1}\vec{p_I}}{(2\pi)^{d-1}}\frac{\left|\phi_I(\vec{p_I})\right|^2}{2E_I}\nonumber\\*
&\phantom{=}\quad\quad\quad\quad\quad\quad\quad\quad\cdot \left|\mathcal{M}\left(p_I\rightarrow p_1\cdots p_n\right)\right|^2(2\pi)^d\delta^{d}\left(p_I-\sum_i p_i\right).
\end{align}

For the final step, we use the fact that we started with an initial state representing a particle at rest to assume that $\phi_I(p_I)$ is strongly peaked around $p_I = m_a$, allowing us to approximate
\begin{align}
X &\approx \frac{K}{2m_a g^2} \int \frac{\sd^{d-1}\vec{p_1}}{\left(2\pi\right)^{d-1}}\frac{1}{2E_1}\cdots\int \frac{\sd^{d-1}\vec{p_n}}{\left(2\pi\right)^{d-1}}\frac{1}{2E_n}\nonumber\\*
&\phantom{\approx}\quad\quad\quad\quad\quad\cdot \left|\mathcal{M}\left(m_a\rightarrow p_1\cdots p_n\right)\right|^2(2\pi)^d\delta^{d}\left(p_I-\sum_i p_i\right).
\end{align}
and hence $X \approx \frac{K}{g^2}\Gamma_{\text{QFT}}$. 

Since both the Unruh detector decay rate and the decay rate calculated through QFT are proportional to $X$, they are necessarily proportional to each other. Since both contain a coupling strength, these can be equated such that both models give the same result in identical situations.

\section{Example: Real scalar field decay $\phi_1 \rightarrow \phi_2\phi_3$}\label{sec:example}

We can check this equivalence with an explicit calculation in the case of the decay $\phi_1 \rightarrow \phi_2\phi_3$, where all fields are real and scalar. We will see that the results indeed match up when taking the appropriate limits.

\subsection{Rate from the Unruh detector perspective}

For the Unruh detector decay rate, we will map the $\phi_1$ field to the $\ket{1}$ state, and the $\phi_2$ state to the $\ket{0}$ state. For the detector operator $M$ we use $\ket{1}\bra{0}+\ket{0}\bra{1}$.

Given this mapping, we only keep the $\phi_3$ field around as an actual field. Let us begin by calculating the two-point function for this field. A straightforward calculation gives
\begin{align}
g_3(t) &\phantom{=\joinrel=\joinrel=\joinrel}=\phantom{\joinrel=\joinrel=\joinrel=} \bra{\psi_{3,0}}\phi_3((0,0,\ldots,0))\phi_3((t,0,\ldots,0))\ket{\psi_{3,0}}\\
&\phantom{=\joinrel=\joinrel=\joinrel}=\phantom{\joinrel=\joinrel=\joinrel=} \frac{1}{2}\int\frac{\sd^{d-1}\vec{p}}{\left(2\pi\right)^{d-1}}\frac{e^{it\sqrt{\vec{p}^2+m_3^2}}}{\sqrt{\vec{p}^2+m_3^2}}\\
&\overset{p=\left|\vec{p}\right|}{\phantom{=\joinrel=\joinrel}=\joinrel=\joinrel=\phantom{\joinrel=\joinrel=}} \frac{\pi^{\frac{d-1}{2}}}{\Gamma\left(\frac{d-1}{2}\right)}\int\limits_0^\infty\frac{\sd p}{\left(2\pi\right)^{d-1}}p^{d-2}\frac{e^{it\sqrt{p^2+m_3^2}}}{\sqrt{p^2+m_3^2}}\\
&\overset{u=\sqrt{1+\frac{p^2}{m_3^2}}}{=\joinrel=\joinrel=\joinrel=\joinrel=\joinrel=\joinrel=} \frac{m_3^{d-2}}{2^{d-1}\pi^{\frac{d-1}{2}}\Gamma\left(\frac{d-1}{2}\right)}\int\limits_1^\infty\sd u \frac{e^{im_3ut}}{\left(u^2-1\right)^{\frac{3-d}{2}}}\\
&\phantom{=\joinrel=\joinrel=\joinrel}=\phantom{\joinrel=\joinrel=\joinrel=} -\frac{1}{4}\left(\frac{m_3}{2\pi\left|t\right|}\right)^{\frac{d-2}{2}}\left(Y_{\frac{2-d}{2}}\left(m_3\left|t\right|\right)-i\sgn(t)J_{\frac{2-d}{2}}\left(m_3\left|t\right|\right)\right)
\end{align}
assuming $1 < d < 3$. The last step follows from formula~10.9.12~in~\cite{DLMF}.

Using this, we can calculate the decay rate from Equation~\ref{eq:decay_unruh}
\begin{align}
\Gamma &= G^2\int\limits_{-\infty}^\infty\sd \kappa e^{-i\Delta\kappa}g_3(\kappa)\\
&= -G^2\int\limits_{-\infty}^\infty\sd \kappa\left(\cos\left(\Delta\left|\kappa\right|\right)-i\sgn(\kappa)\sin\left(\Delta\left|\kappa\right|\right)\right)\nonumber\\*
&\phantom{=}\quad\quad\cdot\frac{1}{4}\left(\frac{m_3}{2\pi\left|\kappa\right|}\right)^{\frac{d-2}{2}}\left(Y_{\frac{2-d}{2}}\left(m_3\left|\kappa\right|\right)-i\sgn(\kappa)J_{\frac{2-d}{2}}\left(m_3\left|\kappa\right|\right)\right)\\
&= \frac{G^2}{2}\left(\frac{m_3}{2\pi}\right)^{\frac{d-2}{2}}\int\limits_0^{\infty}\sd\kappa\quad\kappa^{\frac{2-d}{2}}\sin\left(\Delta\kappa\right)J_{\frac{2-d}{2}}\left(m_3\kappa\right)\nonumber\\*
&\phantom{=} -\frac{G^2}{2}\left(\frac{m_3}{2\pi}\right)^{\frac{d-2}{2}}\int\limits_0^{\infty}\sd\kappa\quad\kappa^{\frac{2-d}{2}}\cos\left(\Delta\kappa\right)Y_{\frac{2-d}{2}}\left(m_3\kappa\right)
\end{align}

This last line can be evaluated using equations~6.699.5 and~6.699.14 in~\cite{GradRyzh}. Using these, we find
\begin{align}
\Gamma = \begin{cases}
G^2\frac{2^{2-d}\pi^{\frac{3-d}{2}}}{\Gamma\left(\frac{d-1}{2}\right)}\left(\Delta^2-m_3^2\right)^{\frac{d-3}{2}} & \text{if}\quad \Delta > m_3, \\
0 & \text{otherwise}.
\end{cases}
\end{align}

As derived, this only holds for $1 < d < 3$. However, it is well defined for all $d > 1$ and analytic in $d$, so by the procedure of dimensional regularization we can assume this to be the decay rate for any dimension $d$ we are interested in.

\subsection{Rate from the QFT perspective}

Comparatively the calculation from the QFT perspective is a bit more straightforward:
\begin{align}
\Gamma &\phantom{=\joinrel}=\phantom{\joinrel=} \frac{g^2}{2m_1}\int\frac{\sd^{d-1}\vec{p_2}}{\left(2\pi\right)^{d-1}}\frac{1}{2\sqrt{\vec{p_2}^2+m_2^2}}\int\frac{\sd^{d-1}\vec{p_3}}{\left(2\pi\right)^{d-1}}\frac{1}{2\sqrt{\vec{p_3}^2+m_3^2}}\nonumber\\*
&\phantom{=\joinrel=\joinrel=}\quad\quad\quad\quad\quad\quad\cdot \left(2\pi\right)^d\delta^d\left(m_1-p_2-p_3\right)\\
&\overset{p=\left|\vec{p_2}\right|}{=\joinrel=\joinrel=}\!\frac{g^2}{m_1}\frac{\pi^{\frac{d-1}{2}}}{\Gamma\left(\frac{d-1}{2}\right)}\int\limits_0^\infty\frac{\sd p}{\left(2\pi\right)^{d-1}}p^{d-2}\frac{\left(2\pi\right)\delta\left(\!m_1\!-\!\sqrt{p^2+m_2^2}-\!\sqrt{p^2+m_3^2}\right)}{4\sqrt{p^2+m_2^2}\sqrt{p^2+m_3^2}}\\
&\phantom{=\joinrel}=\phantom{\joinrel=}\!\!\begin{cases}
\left.\frac{g^2\pi^{\frac{3-d}{2}}p^{d-3}}{2^dm_1\Gamma\left(\frac{d-1}{2}\right)}\frac{1}{\sqrt{p^2+m_2^2}+\sqrt{p^2+m_3^2}}\right|_{p=\frac{\sqrt{\lambda(m_1^2,m_2^2,m_3^2)}}{2m_1}} &\!\!\text{if}\; m_1 > m_2+m_3\\
0 &\!\!\text{otherwise}
\end{cases}
\end{align}

Note that, at first glance, this is not the same as the result from the Unruh detector. However, the Unruh detector model is the result of taking a limit in which both $m_1\rightarrow\infty$ and $m_2\rightarrow\infty$ whilst keeping their difference $\Delta = m_1-m_2$ constant. Taking this limit, whilst also keeping $G = \frac{g}{2m_2}$ constant to keep the overall decay rate from going to zero, we find
\begin{align}
\Gamma &= \lim_{m_2\rightarrow\infty}\left.\frac{2^{2-d}m_2^2G^2\pi^{\frac{3-d}{2}}p^{d-3}}{\left(m_2+\Delta\right)\Gamma\left(\frac{d-1}{2}\right)}\frac{1}{\sqrt{p^2+m_2^2}+\sqrt{p^2+m_3^2}}\right|_{p=\frac{\sqrt{\lambda((m_2+\Delta)^2,m_2^2,m_3^2)}}{2(m_2+\Delta)}}\\
&= \left.G^2\frac{2^{2-d}\pi^{\frac{3-d}{2}}p^{d-3}}{\Gamma\left(\frac{d-1}{2}\right)}\right|_{p=\sqrt{\Delta^2-m_3^2}} = G^2\frac{2^{2-d}\pi^{\frac{3-d}{2}}}{\Gamma\left(\frac{d-1}{2}\right)}\left(\Delta^2-m_3^2\right)^{\frac{d-3}{2}}
\end{align}
which matches the expression obtained from the Unruh detector.

\subsection{Impact of acceleration on the decay rate}

The results from Section~\ref{sec:accelerated_decay} can now be used to calculate the decay rate for non-zero accelerations. We implemented numeric integration using Simpson's rule to evaluate integral~\ref{eq:apply_accel}. This code relies on~\cite{archt} for the calculation of the Bessel functions of imaginary order. All used code is available at \url{https://github.com/davidv1992/fourierdecay}.

\begin{figure}
    \center{\includegraphics[width=0.8\textwidth]{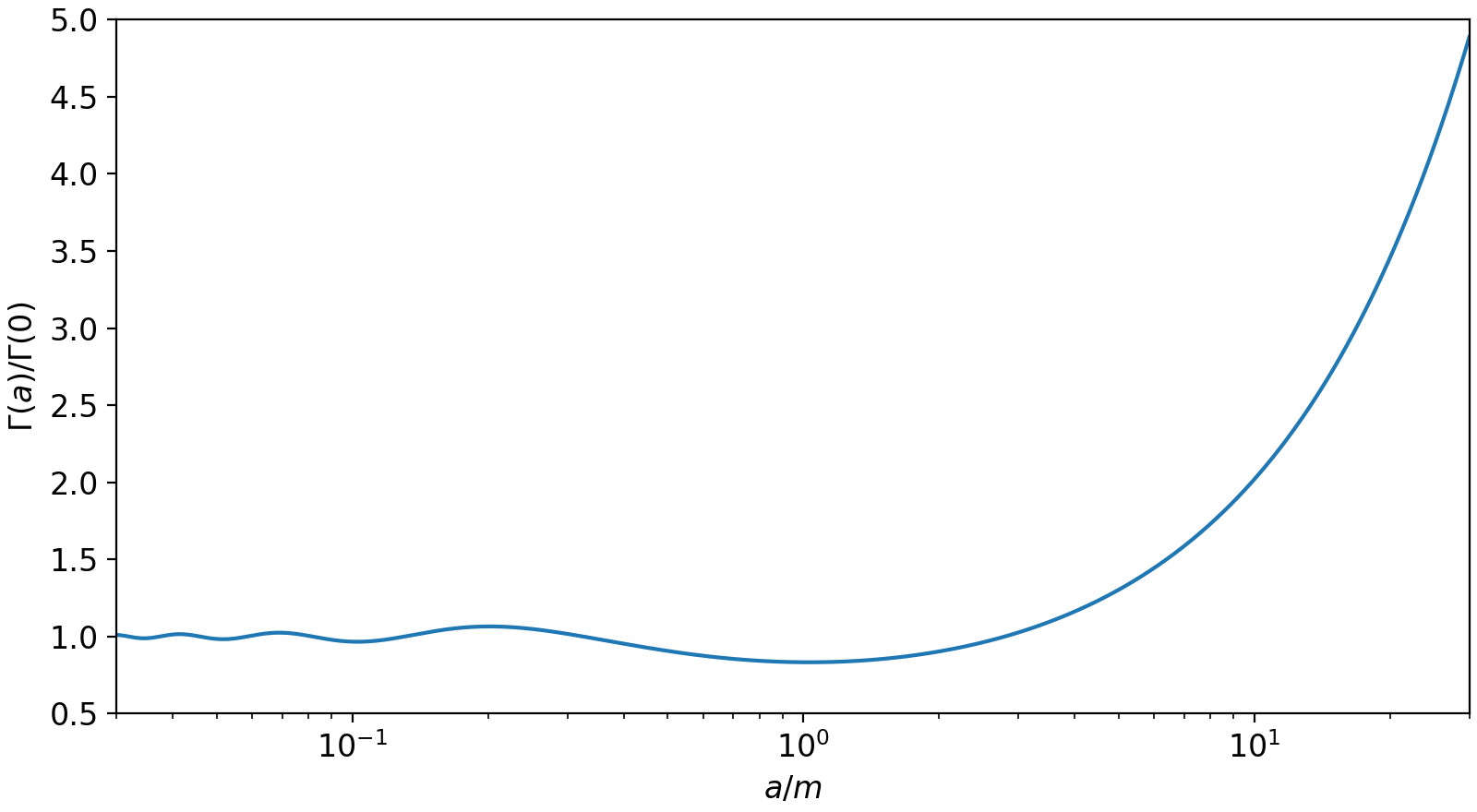}}
    \caption{Decay rate relative to the non-accelerated case for a particle with $\Delta = 1.5m$, in 3+1 dimensions.}
    \label{fig:acceldecay_4d}
\end{figure}

Figure~\ref{fig:acceldecay_4d} shows the change in decay rate as a function of acceleration in a 3+1 dimensional spacetime. We see slight oscillations in the decay rate below $a=m$, with a more significant departure above that boundary. Compared with the 1+1 dimensional case studied in more detail in \cite{self_decay}, the oscillations are damped as $a\rightarrow 0$ and the departure is towards increased decay rather than decreased decay, but otherwise the behaviour looks qualitatively similar.

\begin{figure}
    \center{\includegraphics[width=0.8\textwidth]{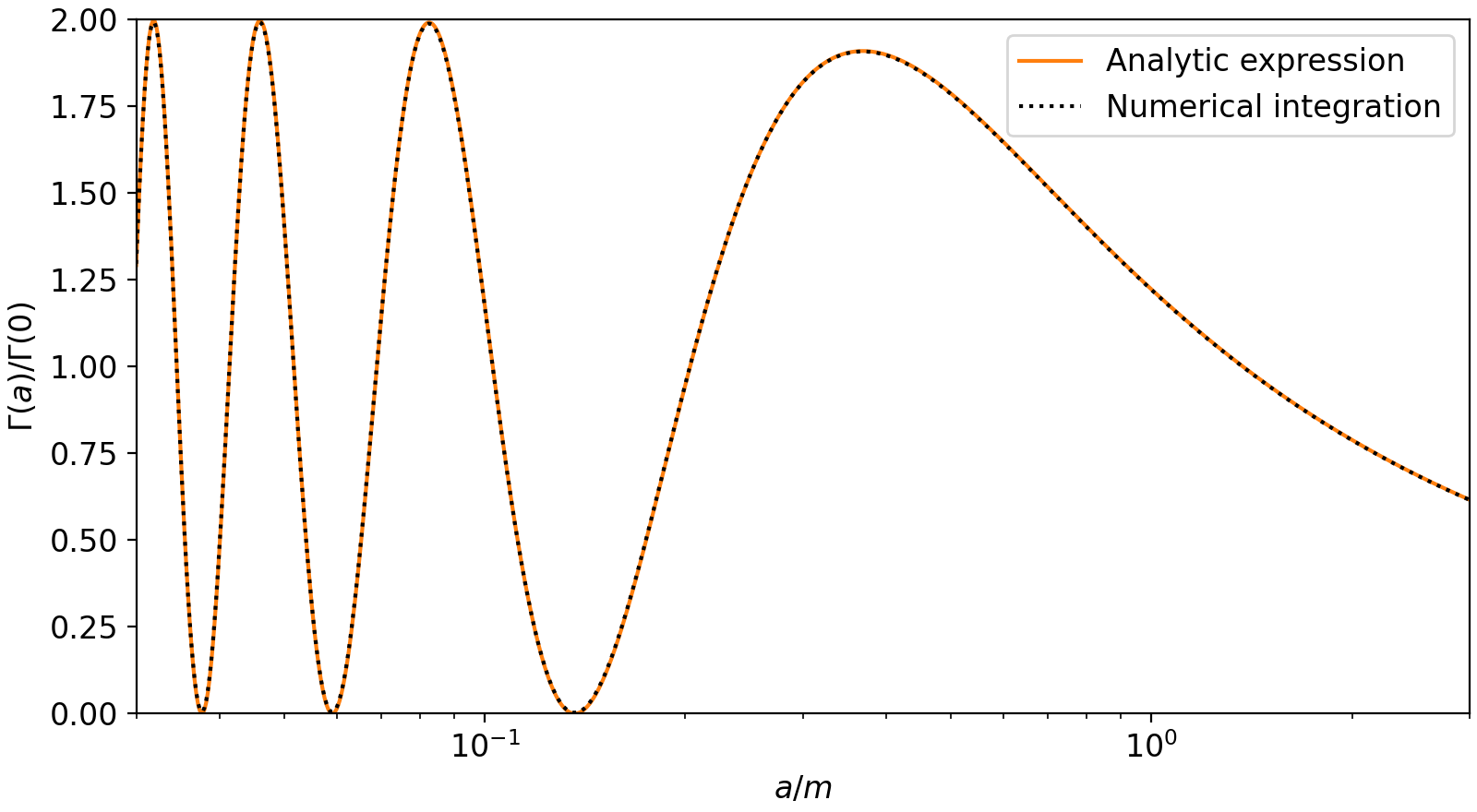}}
    \caption{Decay rate relative to the non-accelerated case for a particle with $\Delta = 1.5m$, in 1+1 dimensions. Both the numerically integrated result (dotted line), as well as the direct evaluation of the closed analytical form (solid line) are shown.}
    \label{fig:acceldecay_2d_check}
\end{figure}

In a 1+1 dimensional spacetime, the accelerated decay rate can be calculated analytically\cite{self_decay}. Figure~\ref{fig:acceldecay_2d_check} shows both the decay rate calculated through numerical integration, as well as the analytic result. We see there is good agreement between the two expressions.

\section{Generalizations}

Both the Unruh detector model from section~\ref{sec:unruhmodel}, as well as the equivalence proof from section~\ref{sec:equivalence} readily generalize to other particle types. We will not write out the full derivations here, but rather focus on the main differences and subtleties.

The most straightforward case is the case where both the initial and the heavy final state particle are scalar. In this scenario, only the particle fields of the Unruh detector get changed, and the proof of equivalence to the QFT decay rate is nearly identical to the cases with only scalar fields.

Additional complications arise when one or both of the particles identified with Unruh states have non-zero spin. Although there are other approaches, we will here only consider the solution of choosing a specific spin state for the identified particles. This yields a slightly modified interaction term of the form
\begin{align}
\overline{f}Fg\ket{0}\bra{1} + h.c.
\end{align}
where $f$ and $g$ are general spin vectors (e.g. a spinor if the particle is spin 1/2, or a polarization vector for spin 1), and $F$ is some product of the other fields. Note that when the spin of the $\ket{0}$ and $\ket{1}$ states are non-identical $F$ is a non-square matrix. The proof of equivalence still follows the form outlined in section~\ref{sec:equivalence}, with the note that care should be taken to only consider the decay with the chosen initial and final spin states when calculating the matrix element.

Finally, let us briefly consider decays with identical particles in the final state. Note that by the assumption that one of the decay products is heavy precludes the option that this is the case for the particle associated with the $\ket{0}$ state of the detector. To deal with such identical particles, only a single field per particle type is used in the Unruh detector model. Rather, the interaction term is modified to include that field multiple times. This results in a $2n$-point function for the field of a particle occurring $n$ times in the final state in the expression for the decay rate of the Unruh model, in place of the 2-point function. For example, for the process $\phi_1\rightarrow \phi_2\phi_3\phi_3$ Equation~\ref{eq:raw_decay_rate} becomes
\begin{align}
\Gamma &= \lim_{T\rightarrow\infty}\frac{G^2}{2T}\left|\bra{0}M\ket{1}\right|^2\int\limits_{-T}^T\sd\tau_1\int\limits_{-T}^T\sd\tau_2 e^{i\Delta(\tau_1-\tau_2)}\nonumber\\*
&\phantom{=}\quad\quad\quad\quad\quad\quad \cdot \bra{\psi_{3,0}}\phi_3(x(\tau_1))\phi_3(x(\tau_1))\phi_3(x(\tau_2))\phi_3(x(\tau_2))\ket{\psi_{3,0}}.
\end{align}
As long as this change to $2n$-point function is properly carried forward, the same technique as in section~\ref{sec:equivalence} can still be used to show the decay rate to be identical between the Unruh model and QFT calculations.

\section{Conclusions and outlook}

Our results provide a significant shortcut in the calculation of the decay rate of accelerated particles. This was achieved through two major tools: First, we found a numerically integrable relation between the transition rate of an accelerated Unruh detector and that of a stationary one. Second, we've shown a relationship between the transition rate of a stationary Unruh detector and an equivalent process in a quantum field theory that can be evaluated through the Feynman diagram formalism. Combined, these make calculating the decay rate of accelerated particles to first order in coupling constants a mostly mechanical calculation exercise.

The form of the relationship found between the accelerated and stationary Unruh detector suggests that changes to the transition rate of the Unruh detector result from a change to the energy available in the transition to excite the fields. This intuitively makes sense as the process that causes acceleration is expected to be able to provide energy to the system. Translating this to the language of Feynman diagrams, it appears that rather than just the decay process diagram itself being involved, there are also diagrams contributing with additional initial state particles coming from the field that is causing the acceleration. Making these links more concrete could provide an interesting avenue of research for further understanding the nature of the Unruh effect.

Furthermore, variations on the Unruh detectors studied here are also used in other applications, such as the study of entanglement. It would be interesting to see whether the results found here can be generalized to also simplify calculations for these areas of study.

Finally, our calculation methods significantly ease the study of more complicated decay processes. As such, it would be interesting to look whether there are decay processes not considered before that are sufficiently low energy such that the effects of acceleration could potentially be studied in a laboratory setting. Predictions for such processes could guide future efforts for experimental observations of the Unruh effect.

\printbibliography

\appendix

\section{Properties of decay rates in $d$ dimensions}\label{ap:square_integrable}

To motivate the applicability of the results in section~\ref{sec:accelerated_decay}, we will show here that the decay rate for an $n$-particle decay is square integrable ($L^2$) for some range of $d$ when using dimensional regularization. This then allows us to use Fourier theory to calculate a decay rate in that range, after which analytic continuation can be used to extend it to values of $d$ of physical interest.

Based on the equivalence shown in section~\ref{sec:equivalence}, we can calculate the decay rate from QFT. This provides a significant shortcut, giving us a starting point of
\begin{align}
\Gamma &= \frac{g^2}{2m_a}\int\frac{\sd^{d-1}\vec{p_n}}{\left(2\pi\right)^{d-1}}\frac{1}{2E_n}\cdots\int\frac{\sd^{d-1}\vec{p_1}}{\left(2\pi\right)^{d-1}}\frac{1}{2E_1}\left(2\pi\right)^d\delta^{d}(m_a-\sum_ip_i)
\end{align}
assuming a point-like interaction and initial particle at rest. We have chosen to reverse the order of labeling the momentum integrals to simplify next steps.

To aid the analysis, we will split this into a recursive formulation:
\begin{align}
\Gamma &= \frac{g^2}{2m_a}X_n(m_a),\\
X_2(p_{T,2}) &= \int\frac{\sd^{d-1}\vec{p_2}}{\left(2\pi\right)^{d-1}}\frac{1}{2E_2}\int\frac{\sd^{d-1}\vec{p_1}}{\left(2\pi\right)^{d-1}}\frac{1}{2E_1}\left(2\pi\right)^d\delta^{d}(p_{T,2}-p_2-p_1),\\
X_i(p_{T,i}) &= \int\frac{\sd^{d-1}\vec{p_i}}{\left(2\pi\right)^{d-1}}\frac{1}{2E_i}X_{i-1}(p_{T,i}-p_i).
\end{align}
Note that the functions $X_i$ are invariant under Lorentz transformation of their argument $p_{T,i}$. Therefore we can use a boost to express $X_i$ as a function of the center of mass energy of $p_{T,i}$. This allows us to simplify the expression by doing every step in the decomposition of the phase space in the center of mass frame for the remaining final-state particles:
\begin{align}
\Gamma &= \begin{cases}
\frac{g^2}{2m_a}Y_n(m_a) & \text{if}\; m_a > \Sigma_n\\
0 & \text{otherwise}
\end{cases},\\
Y_2(E_{T,2}) &= \int\limits_0^\infty \frac{\sd p}{\left(2\pi\right)^{d-1}}\frac{2\pi^{\frac{d-1}{2}}p^{d-2}}{\Gamma\left(\frac{d-1}{2}\right)}\frac{2\pi\delta\left(E_{T,2}-\sqrt{m_1^2+p^2}-\sqrt{m_2^2+p^2}\right)}{4\sqrt{m_1^2+p^2}\sqrt{m_2^2+p^2}}\\
&= \left.\frac{p^{d-3}}{2^{d-1}\pi^{\frac{d-3}{2}}\Gamma\left(\frac{d-1}{2}\right)}\frac{1}{\sqrt{p^2+m_1^2}+\sqrt{p^2+m_2^2}}\right|_{p=\frac{\sqrt{\lambda(E_{T,2}^2,m_1^2,m_2^2)}}{2E_{T,2}}}\\
Y_i(E_{T,i}) &= \int\limits_0^{\frac{\sqrt{\lambda(E_{T,i}^2,\Sigma_{i-1}^2,m_i^2)}}{2E_{T,i}}}\sd p_i\frac{p_i^{d-2}Y_{i-1}\left(\sqrt{\left(E_{T,i}-\sqrt{m_i^2+p_i^2}\right)^2-p_i^2}\right)}{2^{d-1}\pi^{\frac{d-1}{2}}\Gamma\left(\frac{d-1}{2}\right)\sqrt{p_i^2+m_i^2}}
\end{align}
where we define $\Sigma_i = \sum_{j=1}^i m_j$.

We are now in a position to take the simultaneous limit of $m_1$ and $m_a = m_1+\Delta$ to infinity, again keeping $G=\frac{g}{2m_1}$ constant. This further simplifies to our final form
\begin{align}
\Gamma &= \begin{cases}
G^2Z_n(\Delta) & \text{if}\; \Delta > \overline{\Sigma}_n\\
0 & \text{otherwise}
\end{cases},\\
Z_2(\Delta_2) &= \frac{\left(\Delta_2^2-m_2^2\right)^{\frac{d-3}{2}}}{2^{d-2}\pi^{\frac{d-3}{2}}\Gamma\left(\frac{d-1}{2}\right)}\\
Z_i(\Delta_i) &= \int\limits_0^{\sqrt{(\Delta_i-\overline{\Sigma}_i)(\Delta_i-\overline{\Sigma}_i+2m_i)}}\sd p_i \frac{p_i^{d-2}Z_{i-1}\left(\Delta_i-\sqrt{m_i^2+p_i^2}\right)}{2^{d-1}\pi^{\frac{d-1}{2}}\Gamma\left(\frac{d-1}{2}\right)\sqrt{p_i^2+m_i^2}}
\end{align}
where $\overline{\Sigma}_i = \sum_{j=2}^i m_j$. From this formulation, it is clear that $\Gamma$ is $L^2$ but for two complications: A potential pole at $\Delta = \overline{\Sigma}_n$, and its behaviour as $\Delta\rightarrow\infty$.

Both these complications depend entirely on the behaviour of $Z_n$, and neither depends on the scaling of $Z_n$. Hence, it is sufficient for our purposes to show that the following simplified functions are $L^2$ for some range of $d$:
\begin{align}
\overline{Z}_2(\Delta_2) &= \left(\Delta_2^2-m_2^2\right)^{\frac{d-3}{2}},\\
\overline{Z}_i(\Delta_i) &= \int\limits_0^{\sqrt{(\Delta_i-\overline{\Sigma}_i)(\Delta_i-\overline{\Sigma}_i+2m_i)}}\sd p_i \frac{p_i^{d-2}}{\sqrt{m_i^2+p_i^2}}\overline{Z}_{i-1}\left(\Delta_i - \sqrt{m_i^2+p_i^2}\right).
\end{align}

\subsection{Behaviour near $\Delta=\overline{\Sigma}_n$}.

To study the behaviour near $\Delta=\overline{\Sigma}_n$, i.e. $\Delta_n=\overline{\Sigma}_n$, we introduce $t_i = \Delta_i - \overline{\Sigma}_i$. Reformulating the expressions for $\overline{Z}_i$, we find
\begin{align}
\overline{Z}_2(\overline{\Sigma}_2 + t_2) &= \left(2m_2t_2 + t_2^2\right)^{\frac{d-3}{2}},\\
\overline{Z}_i(\overline{\Sigma}_i + t_i) &= \!\int\limits_0^{\sqrt{2m_it_i + t_i^2}}\!\sd p \frac{p^{d-2}}{\sqrt{m_i^2+p^2}}\overline{Z}_{i-1}(\overline{\Sigma}_{i-1} + t_i+m_i-\sqrt{m_i^2+p^2}).
\end{align}

Assuming all particles are massive, we can show by induction in $n$ that $\overline{Z}_n(\overline{\Sigma}_n + t_n) \approx t_n^{\frac{(n-1)d-(n+1)}{2}}\left(\prod_{i=2}^n\left(2m_i\right)^{\frac{d-3}{2}}\right)\left(\prod_{j=1}^{n-2} B(\frac{d-1}{2},\frac{jd-j}{2})\right)$ when $t_n << m_i$ for all $i$. This can be verified by noting that this is trivially true for $n=2$, and then calculating assuming $n>2$:
\begin{align}
\overline{Z}_n(\overline{\Sigma}_n + t_n) &\approx \int\limits_0^{\sqrt{2m_nt_n}}\sd p\frac{p^{d-2}}{m_n} \overline{Z}_{n-1}(\overline{\Sigma}_{n-1} + t_n-\frac{p^2}{2m_n})\\
&\approx \int\limits_0^{\sqrt{2m_nt_n}}\sd p\frac{p^{d-2}}{m_n} \left(t_n-\frac{p^2}{2m_n}\right)^{\frac{(n-2)d-n}{2}}\left(\prod_{i=2}^{n-1}\left(2m_i\right)^{\frac{d-3}{2}}\right)\nonumber\\*
&\phantom{=}\quad\quad\quad\quad\cdot\left(\prod_{j=1}^{n-3} B(\frac{d-1}{2},\frac{jd-j}{2})\right)\\
&= \int\limits_0^1\sd x x^{\frac{d-3}{2}} \left(1-x\right)^{\frac{(n-2)d-n}{2}}t_n^{\frac{(n-1)d-(n+1)}{2}}\left(\prod_{i=2}^{n}\left(2m_i\right)^{\frac{d-3}{2}}\right)\nonumber\\*
&\phantom{=}\quad\quad\quad\quad\cdot\left(\prod_{j=1}^{n-3} B(\frac{d-1}{2},\frac{jd-j}{2})\right)\\
&= t_n^{\frac{(n-1)d-(n+1)}{2}}\left(\prod_{i=2}^{n}\left(2m_i\right)^{\frac{d-3}{2}}\right)\left(\prod_{j=1}^{n-2} B(\frac{d-1}{2},\frac{jd-j}{2})\right)
\end{align}

Note that this implies that $\overline{Z}_n$ can be integrable near $\Delta_n = \overline{\Sigma}_n$ only when $d>1$, and is $L^2$ only when $d>\frac{n}{n-1}$.

\subsection{Behaviour as $\Delta\rightarrow\infty$}

Before studying the behaviour of $\overline{Z}_i(\Delta_i)$, let us first look at $\overline{Z}_2(\Delta_2)$ specifically. Its form immediately implies that, for any $M > m_2$ there exists a constant $C$ such that $\overline{Z}_2(\Delta_2) < C\Delta_2^{d-3}$ when $\Delta_2 > M$. This, combined with the results shown above, implies that $\overline{Z}_2(\Delta_2)$, and therefore the decay rate of a 2-particle decay, is integrable when $d\in(1,2)$ and $L^2$ when $d\in(2,2.5)$.

We claim that for $\overline{Z}_i(\Delta_i)$ we can find a similar bound when $d<2$. Concretely we will show that for any $M>3\overline{\Sigma}_i$ there exists a $C$ such that $\overline{Z}_i(\Delta_i) < C\Delta_i^{d-3}$ when $\Delta_i > M$. Assume this holds for $i<n$, then
\begin{align}
\overline{Z}_n(\Delta_n) &= \!\!\!\int\limits_0^{\sqrt{(\Delta_n-\overline{\Sigma}_n)(\Delta_n-\overline{\Sigma}_n+2m_n)}}\!\!\sd p_n \frac{p_n^{d-2}}{\sqrt{p_n^2+m_n^2}}\overline{Z}_{n-1}\left(\Delta_n - \sqrt{m_n^2+p_n^2}\right)\\
&= \int\limits_{m_n}^{\Delta_n-\overline{\Sigma}_{n-1}}\sd E \left(E^2-m_n^2\right)^{\frac{d-3}{2}} \overline{Z}_{n-1}\left(\Delta_n - E\right)\\
&= \int\limits_{m_n}^{\frac{\Delta_n}{2}}\sd E \left(E^2-m_n^2\right)^{\frac{d-3}{2}} \overline{Z}_{n-1}\left(\Delta_n - E\right)\nonumber\\*
&\phantom{=}\quad + \int\limits_{\frac{\Delta_n}{2}}^{\Delta_n-\overline{\Sigma}_{n-1}}\sd E \left(E^2-m_n^2\right)^{\frac{d-3}{2}} \overline{Z}_{n-1}\left(\Delta_n - E\right)\\
&\le C_1\Delta_n^{d-3}\int\limits_{m_n}^{\frac{\Delta_n}{2}}\sd E \left(E^2-m_n^2\right)^{\frac{d-3}{2}} + C_2\Delta_n^{d-3}\int\limits_{\overline{\Sigma}_{n-1}}^{\frac{\Delta_n}{2}}\sd E \overline{Z}_{n-1}\left(E\right)\\
&\le C_1\Delta_n^{d-3}\int\limits_{m_n}^{\infty}\sd E \left(E^2-m_n^2\right)^{\frac{d-3}{2}} + C_2\Delta_n^{d-3}\int\limits_{\overline{\Sigma}_{n-1}}^{\infty}\sd E \overline{Z}_{n-1}\left(E\right)
\end{align}
where $C_1$ and $C_2$ are constants whose exact values are irrelevant. The last expression is effectively a constant times $\Delta_n^{d-3}$, as both integrals are finite. Hence by induction the statement holds. Therefore this shows that for an $n$-particle decay the decay rate is an $L^2$ function for $d\in(\frac{n}{n-1},2)$ when $n \ge 3$.
\end{document}